\documentclass[
reprint,
amsmath,
amssymb,
aps,
prl,
]{revtex4-1}

\usepackage{graphicx}
\usepackage{dcolumn}
\usepackage{bm}

\usepackage[colorlinks=true,urlcolor=blue,citecolor=blue,linkcolor=red]{hyperref}
\usepackage{booktabs}
\usepackage{amsmath}
\usepackage{amssymb}
\usepackage{braket}
\usepackage[dvipsnames]{xcolor}
\usepackage{ulem}

\usepackage{amsmath}
\usepackage{graphicx}
\usepackage{ulem}
\usepackage{dcolumn}
\usepackage{epsfig}
\usepackage{bm}
\usepackage{array}
\usepackage{hyperref}
\hypersetup{
colorlinks=true,
citecolor=cyan,
linkcolor=blue,
urlcolor=black,
pdfmenubar=true
}

\usepackage{hyperref}
\usepackage{graphicx}
\usepackage{amsmath}
\usepackage{amsfonts}
\usepackage{amssymb}
\usepackage{xcolor}
\usepackage{bbm}
\usepackage{calrsfs}
\usepackage{dutchcal}
\usepackage{amsthm}

\newcommand{\be}{\begin{equation}}
\newcommand{\ee}{\end{equation}}
\newcommand{\beq}{\begin{eqnarray}}
\newcommand{\eeq}{\end{eqnarray}}

\begin{document}
\title{Non-linear Bragg trap interferometer}
\author{Robin Corgier, Luca Pezzè, and Augusto Smerzi}

\affiliation{
QSTAR, INO-CNR and LENS, Largo Enrico Fermi 2, 50125 Firenze, Italy}

\begin{abstract}
We propose a scheme for trapped atom interferometry 
using an interacting Bose-Einstein condensate. 
The condensate is controlled and spatially split in two confined external momentum modes through a series Bragg pulses. 
The proposed scheme (i) allows the generation of large entanglement in a trapped-interferometer configuration via one-axis twisting dynamic induced by interatomic interaction, and (ii) avoids the suppression of interactions during the interferometer sequence by a careful manipulation of the state before and after phase encoding. 
The interferometer can be used for the measurement of gravity with a sensitivity beyond the standard quantum limit. 
\end{abstract}

\maketitle

{\it Introduction.} Matter-wave 
atom interferometers are ideal tools 
for inertial measurements~\cite{PRL_Canuel_2006,arxiv_Geiger_2020}:
they enable tests of fundamental theories~\cite{ScienceFixler2007,PRLLamporesi2008, PRLGraham2013, NatureRosi2014,PRDChaibi2016,Science_Parker_2018}, as well as practical 
applications such as gravimeters\,\cite{NaturePeters1999, PRADebs2011, NJPAltin2013, NJPFrancis2013, PRLAbend2016}, gradiometers\,\cite{PRLSnadden1998, PRAMcGuirk2002,CQG_Trimeche_2019} and gyroscopes\,\cite{PRLRiehle1991,PRLGustavson1997,NJPGustavson2000,PRLDurfee2006,PRLStockton2011,NJPTackmann2012}.
In the wider context of Grand Unification theory\,\cite{kiefer_quantum_2007}, dual-species matter-wave atom interferometer have been proposed to test in a unique way the weak equivalence principle\,\cite{PRLSchlippert2014,PRLZhou2015,Barrett2016,PRLDuan2016,Asenbaum2020arxiv} where gravity can be tested within a quantum frame-work competing with state-of-the-art classical technologies\,\cite{williams_progress_2004,schlamminger_test_2008,PRLTouboul2017}. 
The use of entangled probe states\,\cite{PRL_Pezze_2009,PRA_Hyllus_2012,PRA_Toth_2012} has been proposed as a viable method to increase the sensitivity of atom interferometers beyond the shot noise limit imposed by uncorrelated-atoms probes\,
\cite{RMPPeeze2018, Nature_Riedel_2010, Nature_Gross_2010, Science_Lucke_2011,
Nature_Hosten_2016, PRL_Kruse_2016, PRL_Sewell_2012, PRL_Braverman_2019, PRL_Malia_2020}.
However, so far, sub-shot noise sensitivities have been mainly shown in proof-of-principle experiments~\cite{RMPPeeze2018} that might not be compatible with the strict experimental conditions imposed by the specific application~\cite{arxiv_Szigeti_2020}. 
For instance, gravimeters require the creation of entangled atoms in controllable and separable momentum modes. 
To generate such states, recent proposals explored the use of high-finesse optical cavities~\cite{PRL_Salvi_2018,PRL_Geiger_2018,QST_shankar_2019} or particle-particle interaction in Bose-Einstein condensates (BECs) where entanglement into internal levels is converted to external degrees of freedom via Raman addressing ~\cite{Nature_Riedel_2010,PRL_Szigeti_2020}. 

\begin{figure*}[t!]
\includegraphics[width=1.0\textwidth]{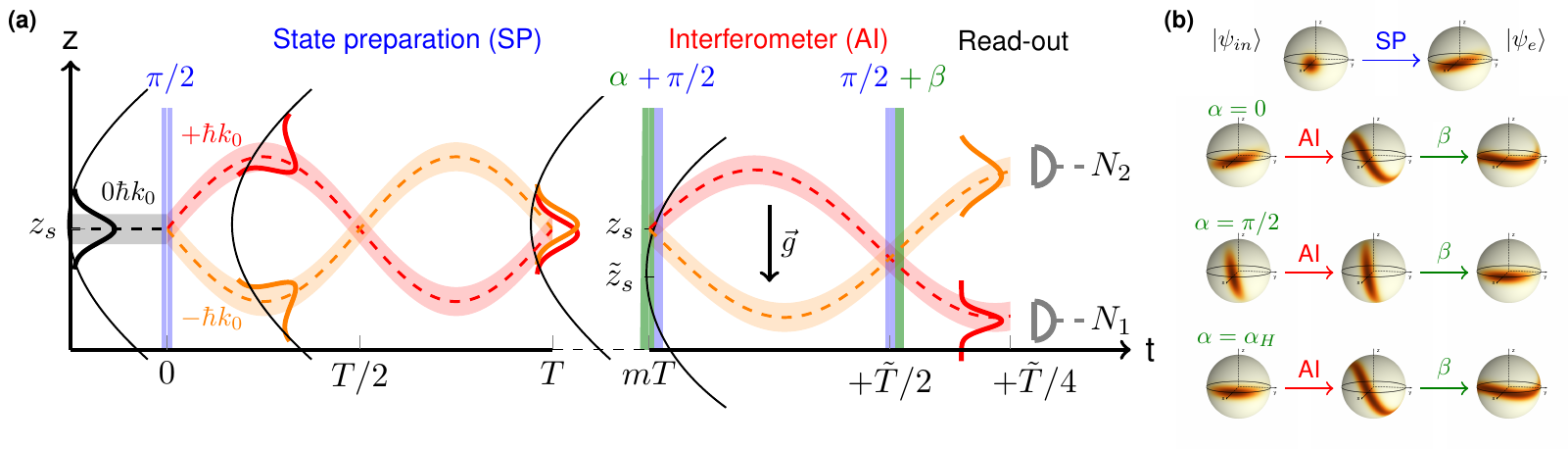}
\caption{
(a) Scheme of the trapped interferometer. At $t=0$, a  $\pi/2$ Bragg pulse (vertical blue pulse) splits the BEC in two momentum modes. The state preparation consists of back-and-forth oscillations up to the time $t=mT$ and is optimized by a final $\alpha$-pulse (vertical green pulse). A $\pi/2$ Bragg pulse starts the interferometer sequence. To be sensitive to gravity, the trap frequency is changed from $\omega_z$ to $\tilde{\omega}_z$. The final $\pi/2$ Bragg pulse closes the interferometer and the state is further optimized by a $\beta$-pulse. The phase is measured by counting the number of atom in each cloud at time $t=mT+3\tilde{T}/4$, when the two output modes are separated and the trap is turn off. (b) The main steps of the sequence are shown in a Husimi Q-distribution on the Bloch sphere. The first row shows the initial coherent spin state at $t=0$ and the generated spin-squeezed state at $t=mT$. The three next rows show the state at different stages of the interferometer: after the $\alpha$ rotation (left column), after phase accumulation (center) and after $\beta$ rotation optimization (right). Here $\alpha_H$ denotes the optimal rotation of the state in the case of a linear interferometer sequence.}
\label{fig_1}
\end{figure*}

In this manuscript, we propose a trapped atom interferometer for the measurement of inertial forces and gravity with a sensitivity beyond the standard quantum limit.
The interferometer uses a trapped interacting BEC with beam-splitters implemented by Bragg pulses~\cite{PRLMuller2008,PRL_Ahlers_2016,PRA_Siemss_2020}, see Fig.~\ref{fig_1}.
Particle entanglement is generated 
in trapped momentum modes via elastic 
atom-atom interaction, which is kept during 
the interferometer sequence.
We show that sub-shot-noise sensitivities can be reached, in our scheme [also referred to as non-linear atom interferometer (NLAI)], thanks to a careful rotation of the state before and after the interferometer sequence that accounts for the growth of phase fluctuations generated by interatomic collisions. 
This avoids the exploitation of a Feshbach resonance to suppress the scattering length between BEC atoms~\cite{Nature_Gross_2010} during the interferometer operations,
which may introduce substantial systematic effects\,\cite{CQGAguilera2014, Hogan08}. 
It thus paves the way toward practical applications of ultra-sensitive trapped-atom interferometry. 

{\it Interferometer scheme.}
The atom interferometer discussed in this manuscript is shown schematically in Fig.~\ref{fig_1}.
It starts with a BEC of $N$ atoms initially at rest in the bottom of a harmonic dipole potential. The trap is kept on during the full interferometer process, until the final readout. 
At $t=0$, a Bragg pulse coherently splits the BEC in two momentum state, $\pm \hbar k_0$, where the effective wave vectors correspond to the two-photon transition $k_0=2k_L$.
Each particle in the BEC is in a quantum superposition of momenta
$\pm \hbar k_0$, such that the state after the Bragg pulse is described by the coherent spin state $|\psi_{in}\rangle= 2^{-S}\sum \limits_{n=0}^{2S}\binom{2S}{n}^{1/2}|S,S-n\rangle$, with $S=N/2$ and the state $|S,S-n\rangle$ indicating $S\pm n$ atoms with momentum $\pm \hbar k_0$, respectively.
Notice that we neglect the possible extra momentum mode generated at each laser pulses \cite{PRA_Hartmann_2020,PRA_Siemss_2020}.
To this aim, different configuration can be considered such as double-Bragg pulses\,\cite{PRL_Ahlers_2016} or the combination of optical lattice and single-Bragg pulses\,\cite{PRLMuller2008,PRA_Siemss_2020}.
Here we assume an infinitely narrow momentum distribution of the input state and justify the use of BEC instead of thermal ensemble\,\cite{NJP_Szigeti_2012}.

The system is described by the field operator $\hat{\Psi}(\mathbf{r},t)= \Phi_{+}(\mathbf{r},t)\,\hat a_{+} +\Phi_{-}(\mathbf{r},t)\,\hat a_{-}$, where $\Phi_{\pm}(\mathbf{r},t)$ carry the spatial evolution of the two wave-function and $\hat{a}_{\pm}$  ($\hat{a}_{\pm}^\dagger$) is the bosonic annihilation (creation) operator of the $\pm$ mode
It is convenient to introduce the SU(2) pseudo-spin operators of the Lie's algebra\,\cite{BookLee1967}, $\hat{S}_x=(\hat{a}^\dagger_{+}\hat{a}_{-}+\hat{a}_{-}^\dagger\hat{a}_{+})/2$, $\hat{S}_y=(\hat{a}_{+}^\dagger\hat{a}_{-}-\hat{a}_{-}^\dagger\hat{a}_{+})/2i$ and $\hat{S}_z =(\hat{a}_{+}^\dagger\hat{a}_{+}-\hat{a}_{-}^\dagger\hat{a}_{-})/2$, satisfying the commutation relation $[S_i,S_j]=i \epsilon_{ijk} S_k$ with $\epsilon_{ijk}$ the Levi-Civita symbol.
Bragg pulses, considered instantaneous at particular time $t_p$, are characterised by an effective Rabi frequency $\Omega_R$ and phase $\phi_L$ and are described by the linear Hamiltonian 
$\hat{H}_0(t)=\hbar \Omega_R\delta(t_p)[\cos(\phi_L)\hat{S}_x+\sin(\phi_L)\hat{S}_y]+\hbar \delta\theta \hat{S}_z$,
where $\delta(t_p)$ is the Dirac delta-function at time $t_p$ and $\delta\theta$ is the precession of the state due to the phase accumulation. Particle-particle interaction is described by 
$\hat{H}_{\mathrm{int}}(t) = \hbar\, \chi(t)\hat{S}_z^2$, where the time dependence in the coefficient $\chi(t)$ is associated to the dynamics of the wave function, see supplementary information~\cite{supp}.

\begin{figure}[t!]
 \includegraphics[width=1.0\columnwidth]{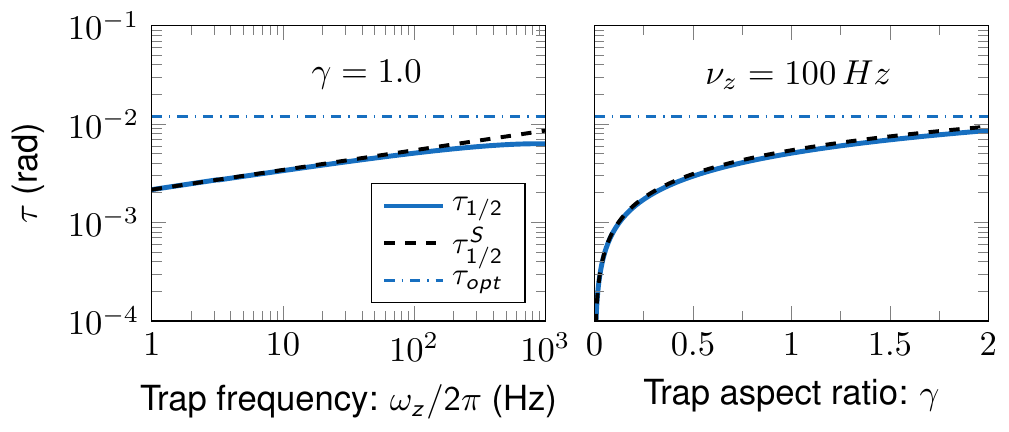}
\caption{Input state preparation. Accumulated non-linear coefficient $\tau_{1/2}$ as a function of $\omega_z$ (left panel) and for spherical trap ($\gamma=1$), and as a function of the trap aspect ration $\gamma$ (right), ranging from pancake geometry ($\gamma<1$) and cylindrical geometry ($\gamma>1$).
The dashed and dot-dashed lines show respectively $\tau_{1/2}$, $\tilde{\tau}$ and $\tau_{opt}$.}
\label{fig_2}
\end{figure}

During the state preparation, no phase is accumulated and the state is described by the one-axis-twisting~\cite{PRA_Kitagawa_1993} transformation 
\be \label{KW}
|\psi_e(mT)\rangle = e^{-i\tau_m \hat{S}_z^2}|\psi_{in}\rangle,
\ee
where $\tau_m=\int_0^{mT} \chi(t)\, dt$ is the accumulated nonlinear coefficient after a time $mT$ depending on the dynamics of the wavepackets $\Phi_{\pm}$ in the trap, $T=2\pi/\omega_z$ and $\omega_z$ the angular trap frequency~\cite{data_description}. 
The calculation of $\tau_m$ can be simplified by neglecting the recombination of the mode at each half period of the trap, giving~\cite{supp}
\begin{equation}
\label{eq_tau_max}
    \tau_{m}^{\rm S}=\dfrac{2\, m\pi}{7}\left(15\,a\,\gamma^2\sqrt{\dfrac{M}{\hbar}}\right)^{2/5}\left(\dfrac{\omega_z}{N^3}\right)^{1/5},
\end{equation}
where $a$ denotes the particle-particle s-wave scattering length, $M$ is the mass of the atom, $\omega_{x,y,z}$ the trap frequencies, $\hbar$ is the Plank constant and $\gamma = \omega_{x,y}/\omega_z$ is the trap aspect ratio.
In practice a fine tuning of $\tau_m$ can be obtained by tuning the trap aspect ratio $\gamma = \omega_{x,y}/\omega_z$ and frequency $\omega_z$. Furthermore, $\tau_{m}^{\rm S}$ linearly increases with the number $m$ of back-and-forth oscillations of the two spatial modes in the trap, see Fig.~\ref{fig_1}.
In Fig.\,\ref{fig_2} we compare $\tau_{m}$ with the approximated $\tau_{m}^{\rm S}$ for the case $m=1/2$, as a function of the trap frequency, panel (a), and trap aspect ratio (b)\,\cite{data_description}.
The entangling evolution (\ref{KW}) can generate a substantial amount of spin squeezing in the state $|\psi_e(mT)\rangle$, which can be quantified by the Wineland parameter $\xi^2 = N( \Delta \hat{S}_z)^2/(\langle \hat{S}_x\rangle^2+\langle \hat{S}_y\rangle^2)$~\cite{PRA_Wineland_1994}.
In particular, the horizontal dot-dashed lines in Fig.\,\ref{fig_2} denotes $\tau_{opt}\approx1.2/N^{2/3}$ leading to the minimum value $\min[\xi(\tau_{opt})] =N^{-1/3}$~\cite{PRL_Pezze_2009}. 
It is important to compare our scheme with that of  Ref.\,\cite{Nature_Riedel_2010}, where entanglement has been generated between a two-component BEC (different internal state of $^{87}$Rb addressed by Raman transitions, see also \cite{PRL_Szigeti_2020}) thanks to a state-dependent potential. Here, entanglement is generated between two different momentum states of a single component BEC (same internal state) where the overlap of the two modes does not inhibit the generation of squeezing.

The interferometer operations consists of two Bragg pulses at time $mT$ and $mT + \tilde{T}/2$, 
described by $R_x(\pi/2)=e^{- i \pi/2 \hat{S}_x}$ and $R_x^\dagger(\pi/2)$, respectively. In between, the angular trap frequency is changed from $\omega_z$ to $\tilde{\omega}_z$, with $\tilde{T}=2\pi/\tilde{\omega}_z$ indicating the period of the new trap, see  Fig.~\ref{fig_1}. 
In this configuration the phase accumulated after half a period is $\theta=2k_0g(1/\tilde{\omega}_z^2-1/\omega_z^2)$\,\cite{Storey_Cohen_1994,PRL_Li_2014},
which makes the apparatus sensitive to gravity. 
Furthermore, the initial (final) pulse can be combined with a rotation of angles $\alpha$ ($\beta$) in order to optimize the state (see discussion below and Fig.\,\ref{fig_1}b).
Using $R_x^\dagger(\pi/2) \, e^{-i (\tilde{\tau}  \hat{S}_z^2+ \theta\hat{S}_z)} \, R_x(\pi/2) =  e^{-i( \tilde{\tau}  \hat{S}_y^2+ \theta\hat{S}_y)}$, the final state of the full interferometer sequence is described by the transformation
\be
\label{eq_out_state}
\vert \psi_f \rangle = e^{-i \beta \hat{S}_x} e^{-i( \tilde{\tau}  \hat{S}_y^2+ \theta\hat{S}_y)} e^{-i \alpha \hat{S}_x}|\psi_e(mT)\rangle
\ee
where $\tilde{\tau}\neq0$ denotes the accumulated extra non-linear coefficient during the interferometry sequence. 

The nonlinear parameters $\tau_m$ (for state preparation) and $\tilde{\tau}$ (for the interferometer sequence) can be tuned independently from each other. 
Notice that the current trapped interferometer sequence avoids the characteristic refocusing $\pi$-pulse of free-falling atom interferometers: the refocusing is provided by the trap geometry. 
At the end of the interferometer sequence, the BEC is kept in the trap for an extra time $\tilde{T}/4$ that guarantees maximum separation between the wave packets. After that, the BEC is released from the trap and imaged.
The phase is estimated by inverting the sinusoidal relation between $\theta$ and the average relative number of particles at the output ports.
In the following we study the sensitivity gain, $\mathcal{G}_{\alpha,\beta}$, of the nonlinear atom interferometer sequence of Fig.\,\ref{fig_1} with respect to the shot noise limit $\Delta \theta_{SN} = 1/\sqrt{N}$. Here $\Delta \theta=\Delta \theta_{SN}/\mathcal{G}_{\alpha,\beta}$ is the phase sensitivity of the NLAI, where $(\Delta \theta)^2 = (\Delta S_z)^2_{\rm \alpha,\beta}/( d\langle S_z \rangle_{\rm \alpha,\beta} / d \theta )^2$ is obtained by error propagation and the spin moments are calculated for the output state of Eq. (\ref{eq_out_state}).
In particular, for $\theta = 0$, we have
\be
\label{eq_gain_def}
\mathcal{G}_{\alpha,\beta}^2=\dfrac{\langle \hat{S}_x\rangle^2_{\rm \alpha,\beta}}{N (\Delta \hat{S}_z)^2_{\rm \alpha,\beta}}\cos^2\beta.
\ee
The indices $\alpha$ and $\beta$ refer to the rotation of the state on the Bloch sphere before and after the interferometer sequence, see Eq.~(\ref{eq_out_state}). In the following, the parameter $\beta$ is always optimized to maximize the sensitivity gain while we consider different choices of $\alpha$: $\alpha=0, \pi/2, \alpha_H$ and $\alpha_{opt}$. The previous first two choices lead, respectively, to a total $\pi/2$ and $\pi$-pulse at time $t=mT$ and can easily be implemented experimentally. The case $\alpha=\alpha_H$ refers to the optimal rotation in the case of a linear interferometer sequence and $\alpha=\alpha_{opt}$ refer to a state protection strategy used to limit the impact of $\tilde{\tau}$ during the interferometry sequence. In the following we denote by $\mathcal{G}_{\alpha_H}^L=\min[\xi(\tau)]$ the sensitivity gain of a linear interferometer ($\tilde\tau=0$).

{\it Weak NLAI.} We first study analytically the situation where $\tilde{\tau}$ is small enough to approximate  $e^{-i\tilde{\tau}  \hat{S}_y^2}\approx1-i\tilde{\tau}\hat{S}_y^2$. 
For $\theta = 0$, we can rewrite Eq.\,(\ref{eq_gain_def}) as~\cite{supp}
\be \label{eq_weal_NLAI}
\mathcal{G}_{\alpha,\beta}^2=\left[1+(2S-1) \{\sin(2\beta)\tilde\tau-\sin\left(2(\alpha+\beta)\right)\tau\}\right]\cos^2\beta.
\ee
In the case $\alpha=0$, the non-linear evolution during the interferometer sequence degrades the sensitivity gain by ``un-squeezing'' the state though the term $\sin(2\beta)(\tilde\tau-\tau)$. 
On the contrary, for $\alpha=\pi/2$, the contribution of $\tilde\tau$ adds to $\tau$ such that the non-linear evolution during the interferometer can  ``squeeze'' the state even further though the term $\sin(2\beta)(\tilde\tau+\tau)$. In this configuration the total pulse at time $mT$ read as $R_x(\pi)$ where the input spin squeezed state is rotated by $\pi$ around $\hat{S}_x$. The total amount of non-linearities can then be simply described by $e^{-i(\tau+\tilde{\tau})\hat{S}_z^2}$ where the $\pi$-pulse at time $t=mT$ does not change the orientation of the spin-squeezed state on the Bloch sphere. Nevertheless in both cases the sensitivity gain, $\mathcal{G}_{\alpha,\beta}$, is strongly impacted in general for $\beta\neq0$ though the term $\cos^2(\beta)$ of Eq.\,(\ref{eq_gain_def}). 
In the case $\beta=0$, the non-linear evolution during the interferometer does not play a role and for $\pi/2 < \alpha < \pi$ the sensitivity is sub-shot-noise. This configuration is equivalent to a linear interferometry sequence optimized for $\alpha_H=-\pi/4$.
The fact that the different rotations do not lead to the same maximum gain emphasizes the importance of a careful pre-rotation and post-rotation of the state to reach $\mathcal{G}>1$. 

\begin{figure}[t!]
\includegraphics[width=1.0\columnwidth]{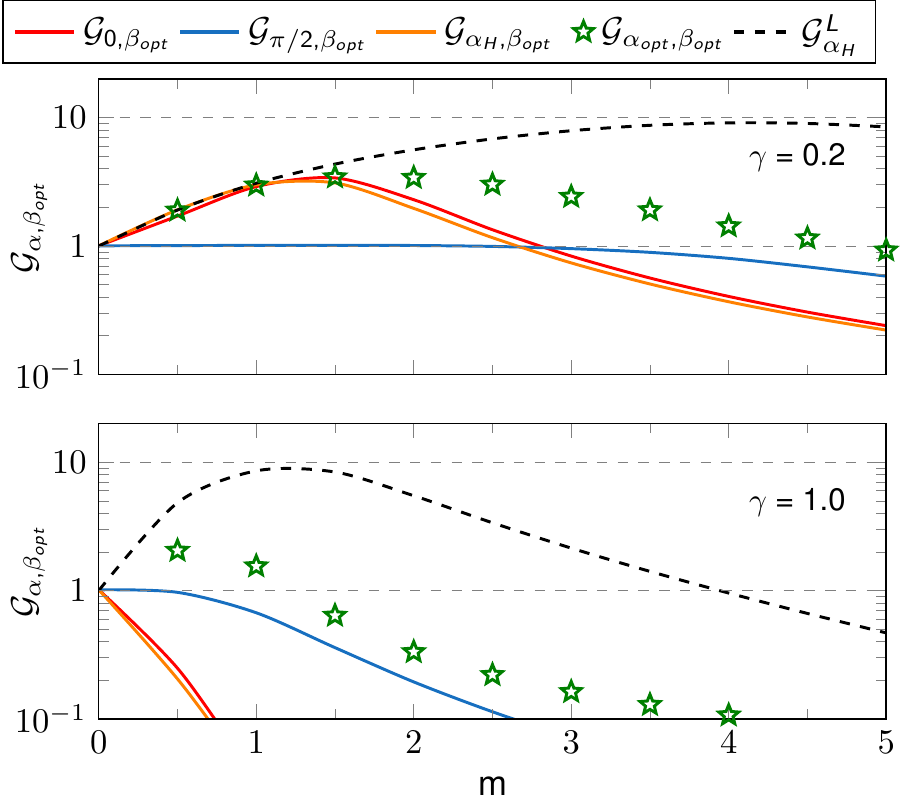}
\caption{Gain factor $\mathcal{G}_{\alpha,\beta_{opt}}$ as a function of $m$ (namely for different spin-squeezed input states). The different panels correspond to different values of the trap aspect ratio $\gamma=0.2$ (pancake shape), $\gamma=1$, (spherical shape), while the different lines refer to a linear interferometer (black dashed) and non-linear interferometer sequence (colored) with $\alpha=0$ (red), $\pi/2$ (blue), $\alpha_H$ (orange) and $\alpha_{opt}$ (green stars). Typically an $\alpha_H$ rotation is used as in the case of linear interferometer and here $\alpha_{opt}$ highlight the optimal rotation of the input state. The numerical  results have been calculated for $m=0,1/2,1...$ and curves are a guide to the eye obtained via spline interpolation. The maximum gain, $\mathcal{G}=N^{1/3}$, and shot-noise limit, $\mathcal{G}=1$, are highlighted by the horizontal dashed lines.}
\label{fig_4}
\end{figure}

{\it Strong NLAI.}
We now study numerically the more realistic case where the non-linear terms of the state preparation and interferometer sequence is not small. Figure\,\ref{fig_4} shows the sensitivity gain as a function of the number of back-and-forth oscillations, $m$, and for different trap aspect ratios. Let us first discuss the case of small interaction, namely $\gamma$ and $m$ small and $\tau<\tau_{opt}$. In the case of $\alpha=\alpha_H$, the results confirm qualitatively the analysis discussed above where $\mathcal{G}_{\alpha_H,\beta_{opt}}\approx\mathcal{G}^L_{\alpha_{H}}$. 
It is interesting to notice that the case $\alpha=0$ gives results very similar to a numerical optimizations over both $\alpha$ and $\beta$ (green stars): $\mathcal{G}_{0,\beta_{opt}}\approx\mathcal{G}_{\alpha_{H},\beta_{opt}}\approx\mathcal{G}^L_{\alpha_{opt}}$. This configuration correspond indeed to an ``effortless strategy'' where only the strength of the final Bragg pulse, closing the interferometer, need to be scanned (and optimized). For $\alpha=0$ the state is given by: $\vert \psi_f \rangle = e^{-i \beta \hat{S}_x} e^{-i \tilde{\tau} \hat{S}_y^2}e^{-i\tau \hat{S}_z^2}|\psi_{in}\rangle$ where the sequential action of first $e^{-i\tau \hat{S}_z^2}$ and then $e^{-i \tilde{\tau} \hat{S}_y^2}$ on the state shears the ellipsoid in two different directions leading to the ``S-shape'' highlighted in Fig.\,\ref{fig_1}b on the Bloch sphere.
In the case where $\alpha=\pi/2$ the state read as $\vert \psi_f \rangle = e^{-i (\beta+\pi/2) \hat{S}_x} e^{-i( \tilde{\tau}+\tau)  \hat{S}_z^2}|\psi_{in}\rangle$. As shown in Fig\,\ref{fig_1}b, in this case the ellipsoid is not deformed and does not exhibit an ``S-shape''. Even though the orientation of the state can be optimized though the $\beta$ rotation, the sensitivity of the interferometer is strongly degraded for $\beta\neq 0$, see Eq.~(\ref{eq_gain_def}).
This explains why the sensitivity is at best shot-noise limited (blue curve).

In the case of strong interaction, namely $\gamma$ and $m$ large, the non-linear term, $\tilde{\tau}$, dramatically degrade the sensitivity gain. In this configuration a non trivial rotation of the input state is needed ($\alpha\neq0$ and $\alpha\neq\alpha_H$) and a trade-off between the deformation of the ellipsoid, ``S-shape'', and final rotation, $\beta\neq0$ degrading the sensitivity gain is required to reach a sub-shot noise sensitivity (green stars).

\begin{figure}[t!]
\includegraphics[width=1.0\columnwidth]{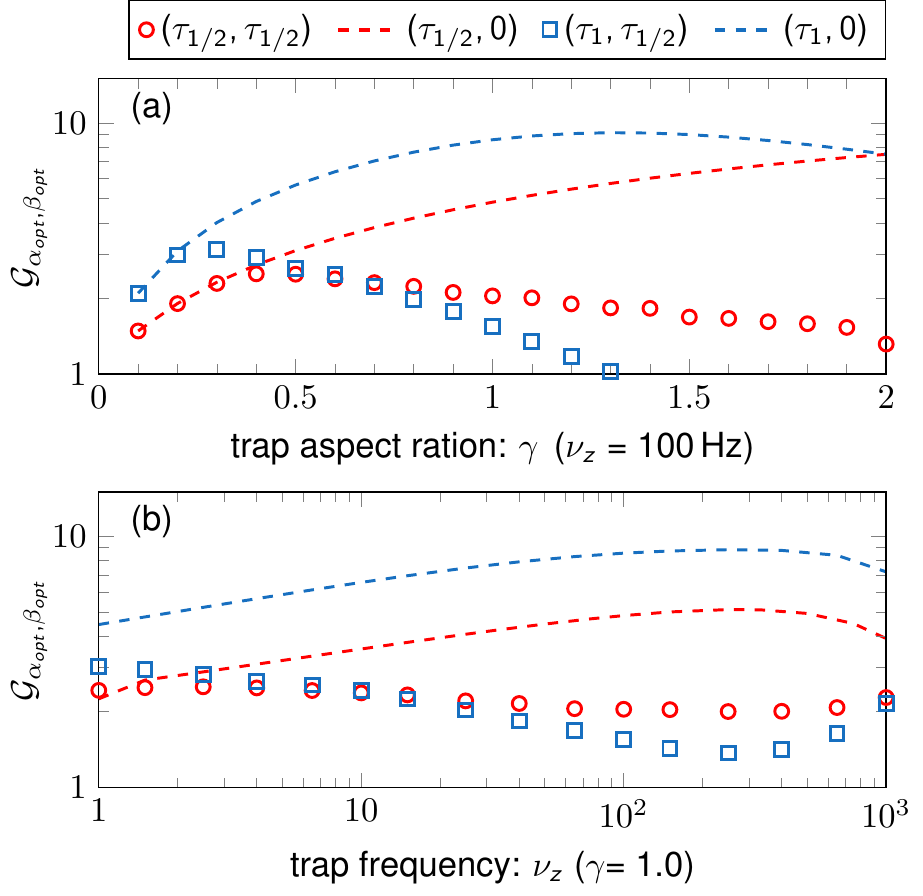}
\caption{Optimized gain factor $\mathcal{G}_{\alpha_{opt},\beta_{opt}}$
as a function of the trap aspect ratio and for a fixed trap frequency $\nu_z=100\,$Hz (a), and as a function of the trap frequency for a spherical traps $\gamma=1$ (b).
Red and blue symbols are obtained for a non-linear interferometer with $m=1/2$ ($\tau=\tau_{1/2}$) and $m=1$  ($\tau=\tau_{1}$), respectively. 
The dashed lines denote the corresponding sensitivity for linear interferometer where $\tilde\tau=0$.}
\label{fig_5}
\end{figure}

Figure \ref{fig_5} shows the sensitivity gain optimized with respect to $\alpha$ and $\beta$ for different trap aspect ratio (panel a)  and different trap frequency (panel b) in the case where the input spin squeezed state is generated by $m=1/2$ (red) or $m=1$ (blue) back-and-forth oscillations. In both case, increasing the number of back-and-forth oscillations benefit to the maximum sensitivity gain where the non-linear terms are controlled though the different traps. Indeed, even so $\tau_{1/2}$ is small, after $m$ back-and-forth oscillations $\tau_m=2m\tau_{1/2}$ while $\tilde\tau=\tau_{1/2}$.
The oscillations of the maximum sensitivity gain, highlighted in panel b, are a direct consequence of the trade-off discuss above. In the case of high trap frequencies and  ``over-squeezed'' input spin squeezed state, $\tau>\tau_{opt}$, the non-linear term, $\tilde\tau$, can  ``un-squeezed'' the state explaining the sudden increase of the optimized sensitivity gain at high trap frequency observed in panel b. 

{\it Discussion and Conclusions.} 
Above, we have assumed a perfectly-harmonic trap configuration. Indeed non-harmonic traps (magnetic or optic) do not prohibit the two modes to overlap classically but could induce a non-identical shape deformation of each mode limiting therefore the efficiency of the different Bragg pulses. Nevertheless, one can expect that the different back-and-forth oscillation would impact the shape of each mode in a similar manner on average limiting the detrimental effect of a shape deformation. In addition, in the case of dipole trap configuration large harmonic traps can be accessible though the ``painted potential'' technique at the cost of laser power\,\cite{Henderson2009,PRARoy2016}.

We have also considered a constant number of atoms while, 
in practice, the shot-to-shot fluctuation in atom number between two consecutive runs can not be avoided. Such fluctuations can be dramatics in regards of the full optimization of the interferometer sequence where the pre- and post-rotation of the state is directly linked to the number of atoms. Though this paper we have shown in the regime of weak-interactions ($N=10^3$ atoms in a $2\pi \{20,20,100\}$\,Hz trap), that the pre-rotation of the state can be avoided ($\alpha=0$) to reach the best sensitivity gain, $\mathcal{G}_{0,\beta_{opt}}=3.5$. The strategy proposed is then consistent with the current technology development and feasibility in current lab experiments\,\cite{Nature_Riedel_2010} where only the final beamsplitter has to be optimized to exhibit sub-shot noise sensitivity measurements.

The trapped atom interferometer proposed in this paper reaches sub-shot noise sensitivities without requiring the suppression of the particle-particle scattering length during phase encoding.
The presence of non-linearity during the interferometer operations can be mitigated via optimal rotations of the state on the Bloch sphere.
Our result are supported by analytical calculations in the regime of small interaction and numerically in the regime of strong interactions. 
Larger trap aspect ratios and/or weaker trap frequency accessible today in micro-gravity environment\,\cite{NatureBecker2018,Nature_Aveline_20210} would benefit to the propose interferometer where arbitrary input spin-squeezed state could be made available but not impacted by a non-linear interferometer sequence.

{\it Acknowledgments.}
The authors thanks Carsten Klempt, Naceur Gaaloul, Samuel Nolan, Francesco Minardi, Chiara Fort, Alessia Burchianti, Chiara d'Errico and Marco Fattori for fruitful discussion.
This work is supported by the European Unions Horizon 2020 research and innovation programme - Qombs Project, FET Flagship on Quantum Technologies grant no. 820419, and from the H2020 QuantERA ERA-NET Cofund in Quantum Technologies projects TAIOL.

\appendix
\onecolumngrid

\section{Time evolution of the Hamiltonian}
\label{sec_Appendix_A}

As introduced in the main text, we consider the two mode field operator $\hat{\Psi}(\mathbf{r},t)= \Phi_{+}(\mathbf{r},t)\,\hat a_{+} +\Phi_{-}(\mathbf{r},t)\,\hat a_{-}$, where $\Phi_{\pm}(\mathbf{r},t)$ carry the spatial evolution of the two wave-functions and $\hat{a}_{\pm}$  ($\hat{a}_{\pm}^\dagger$) is the bosonic annihilation (creation) operator of the $\pm$ mode and obey the usual bosonic commutation rules: $[a_i,a_j^\dagger]=\delta_{i,j}$ and $[a_i,a_i]=[a_i^\dagger,a_i^\dagger]=0$.

\subsection{Linear Hamiltonian}

The linear Hamiltonian explicitly read
\begin{equation} \label{eq_Hamiltonian_L}
    \hat{H}_\mathrm{0}(t)=\int_{-\infty}^{\infty}d\mathbf{r}\, \hat{\Psi}^\dagger(\mathbf{r},t)[H_{\mathrm{K}}+H_{\mathrm{P}} +H_{\rm B}]\hat{\Psi}(\mathbf{r},t)
\end{equation}
where $H_{\mathrm{K}}$ is the kinetic Hamiltonian, $H_{\mathrm{P}}$ the total external potential Hamiltonian, $H_{\mathrm{P}} = M \omega_x^2 x^2/2+M \omega_y^2 y^2/2+M \omega_z^2 (z-z_s)^2$/2, M being the mass of the atom, $\omega_i$ the angular trap frequency in the direction $i=\{x,y,z\}$ and $z_s=-g/\omega_z^2$ denotes the gravitational sag with $g$ the gravity constant. Though the paper the effect of non-harmonic potential are neglected due the possibility to generate wide harmonic potential with painted potential at the expense of laser power\,\cite{Henderson2009,PRARoy2016}. $H_{\rm B}$ denotes the Bragg potential Hamiltonian  characterized though its effective Rabi frequency, $ \Omega_R$ and phase $\phi_L$. General treatment of Bragg pulse can be found in\,\cite{PRLMuller2008,PRA_Siemss_2020}. Introducing the SU(2) pseudo-spin operators of the Lie's algebra\,\cite{BookLee1967}
\begin{eqnarray}
\label{eq_Sx_Sy_Sz_def}
    \hat{S}_x &=& \left(\hat{a}^\dagger_{+}\hat{a}_{-}+\hat{a}_{-}^\dagger\hat{a}_{+}\right)/2, \nonumber \\
    \hat{S}_y &=& \left(\hat{a}_{+}^\dagger\hat{a}_{-}-\hat{a}_{-}^\dagger\hat{a}_{+}\right)/2i,\\  \nonumber
    \hat{S}_z &=&\left(\hat{a}_{+}^\dagger\hat{a}_{+}-\hat{a}_{-}^\dagger\hat{a}_{-}\right)/2,
\end{eqnarray}
satisfying the commutation relation $[S_i,S_j]=i \epsilon_{ijk} S_k$ with $\epsilon_{ijk}$ the Levi-Civita symbol, it is convenient to decompose $\hat{H}_\mathrm{0}(t)$ as 
\be
\hat{H}_0(t)=\hbar \Omega_R\delta(t_p)[\cos(\phi_L)\hat{S}_x+\sin(\phi_L)\hat{S}_y]+\hbar \delta\theta(t) \hat{S}_z,
\ee
where the Bragg-pulse (first two terms) have been considered instantaneous at particular time $t_p$ with $\delta$ the Dirac delta-function. Depending on the choice of the laser phase, rotation of the state can be performed around $\hat{S}_x$ ($\phi_L=0$), $\hat{S}_y$ ($\phi_L=\pi/2$) or a combination of both by an angle defined by the strength of the laser pulse. A $\pi/2$-pulses at time $t_p$ for instance is defined by $\int dt' \Omega_R\delta(t_p-t')=\pi/2\,\delta(t_p)$. Here the contribution of $H_K+H_P$ to $\hat{S}_x$ and $\hat{S}_y$ have been neglected when the two modes overlap. The phase accumulated between two consecutive pulses at time $t_1$ and $t_2$ read\,\cite{Storey_Cohen_1994,PRL_Li_2014}
\be
\theta=\int_{t_1}^{t_2}\,dt\,\delta\theta(t)= 2 k_0 (z(t_2)-z(t_1)),
\ee
where $z(t_1)$ and $z(t_2)$ denote the overlap position of the two momentum state. In the case where the trap frequency is constant we find $z(t_2)=z(t_1)$ and no phase due to gravity is accumulated.

\subsection{Non-linear Hamiltonian}

The non-linear Hamiltonian describes particle-particle elastic collision explicitly and read
\begin{equation}
\label{eq_Hamiltonian_NL}
    \hat{H}_{\mathrm{int}}(t)=
    \frac{2 \pi \hbar^2 a}{M}\int_{-\infty}^{\infty}d\mathbf{r}\, \left(\hat{\Psi}^\dagger(\mathbf{r},t)\right)^2\left(\hat{\Psi}(\mathbf{r},t)\right)^2,
\end{equation}
Restricting to the case of two momentum modes, the main contribution to the non-linear Hamiltonian considered in this study read
\begin{equation}
    \hat{\Psi}^\dagger\hat{\Psi}^\dagger\hat{\Psi}\hat{\Psi}=\Phi_{+}^4\,\hat{a}_+^\dagger\hat{a}_+^\dagger\hat{a}_+\hat{a}_++\Phi_{-}^4\,\hat{a}_-^\dagger\hat{a}_-^\dagger\hat{a}_-\hat{a}_-+4\Phi_{+}^2\Phi_{-}^2\,\hat{a}_-^\dagger\hat{a}_+^\dagger\hat{a}_+\hat{a}_-
\end{equation}
where we recognized the SPM modulation term (sum of the two first term) and the CPM modulation therm (third term). Using the relation,
\begin{equation}
    (\hat{a}^\dagger_+\hat{a}_+\pm \hat{a}^\dagger_-\hat{a}_-)^2 = \hat{a}^\dagger_+\hat{a}^\dagger_+\hat{a}_+\hat{a}_+\,\pm\,2\,\hat{a}^\dagger_+\hat{a}^\dagger_-\hat{a}_+\hat{a}_-+\hat{a}^\dagger_-\hat{a}^\dagger_-\hat{a}_-\hat{a}_-+\hat{N},
\end{equation}
with $\hat{N}=\hat{a}^\dagger_+\hat{a}_++\hat{a}^\dagger_-\hat{a}_-$ we find:
\begin{subequations}
\begin{align}
   \hat{a}^\dagger_+\hat{a}^\dagger_+\hat{a}_+\hat{a}_++\hat{a}^\dagger_-\hat{a}^\dagger_-\hat{a}_-\hat{a}_-&=\dfrac{\hat{N}^2}{2}+\dfrac{1}{2} (\hat{a}^\dagger_+\hat{a}_+- \hat{a}^\dagger_-\hat{a}_-)^2-\hat{N}\propto2\hat{S}_z^2, \\
    4\,\hat{a}^\dagger_+\hat{a}^\dagger_-\hat{a}_+\hat{a}_-&= \hat{N}^2- (\hat{a}^\dagger_+\hat{a}_+- \hat{a}^\dagger_-\hat{a}_-)^2 \propto-4\hat{S}_z^2.
\end{align}    
\end{subequations}
Using the relation $\int_{-\infty}^{\infty}\,d\mathbf{r}\,\left|\Phi_{+}(\mathbf{r},t)\right|^4=\int_{-\infty}^{\infty}\,d\mathbf{r}\,\left|\Phi_{-}(\mathbf{r},t)\right|^4$ and Eq.\,(\ref{eq_Hamiltonian_NL}) we can rewrite $\hat{H}_{\mathrm{int}}(t)$ as
\be
\hat{H}_{\mathrm{int}}(t) = \hbar\,\chi_{\mathrm{S}}(t) \hat{H}_{\mathrm{S}}  -2\,\hbar\,\chi_{\mathrm{C}}(t) \hat{H}_{\mathrm{C}}\equiv \hbar \chi(t) \hat{S}_z^2,
\ee 
where the first term,
\be
\hat{H}_{\mathrm{S}} = \dfrac{1}{2}
(\hat{a}_{+}^\dagger\hat{a}_{+})^2+\dfrac{1}{2}(\hat{a}_{-}^\dagger\hat{a}_{-})^2 
\ee
is a self-phase modulation (SPM) that contains non-linearity that are local in each mode with
\be
\chi_{\mathrm{S}}(t)=
\frac{4 \pi \hbar\,a}{M}
 \int_{-\infty}^{\infty}\,d\mathbf{r}\,\left|\Phi_{\pm}(\mathbf{r},t)\right|^4.
\ee
and the second term,
\be
\hat{H}_{\mathrm{C}} = 2\, \hat{a}_{+}^\dagger \hat{a}_{-}^\dagger\hat{a}_{+}\hat{a}_{-},
\ee
is a cross-phase modulation (CPM) 
which describes correlation between the two mode with
\be
\chi_{\mathrm{C}}(t) =\frac{4 \pi \hbar\,a}{M} \int_{-\infty}^{\infty}\,d\mathbf{r}\,\left|\Phi_{+}(\mathbf{r},t)\right|^2 \left|\Phi_{-}(\mathbf{r},t)\right|^2.
\ee
Finally we find $ \chi(t)=\chi_{\mathrm{S}}(t)-2\,\chi_{\mathrm{C}}(t)$.

\begin{figure}[t!]
\includegraphics[width=0.45\columnwidth]{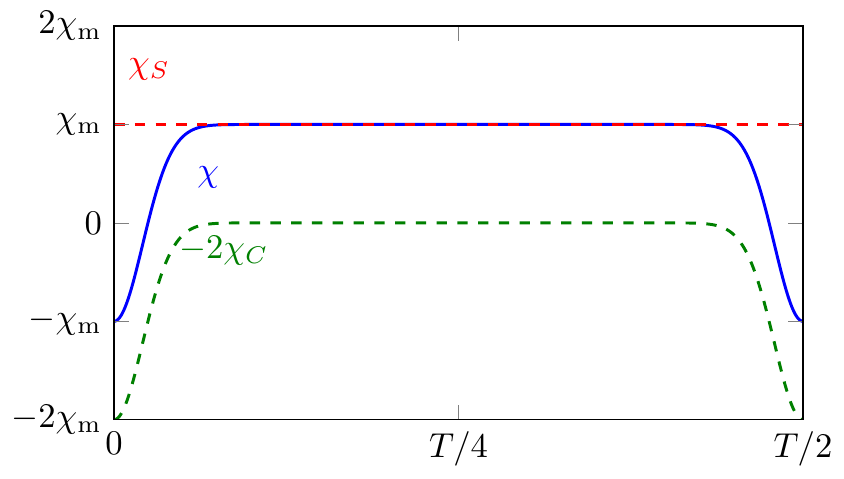}
\caption{Evolution of the non-linear coefficient with time (see Fig.\,\ref{fig_1}).}
\label{fig_6}
\end{figure}

\subsection{Non-linear coefficient}
In the case of small interaction, one can describe the spatial mode though a variational calculation and a Gaussian trial function approach with re-scaled oscillator length. Within this approach the characteristic sizes of the Gaussian $\sigma_i=b\,R_i$ where $R_i$ is the Thomas-Fermi radii in the direction $i\in\{x,y,z\}$, $R_i^2=2 \mu / m\omega_i^2$ with $\mu=\hbar\omega_0/2\left(15Na\sqrt{m\omega_0/\hbar}\right)^{2/5}$ the chemical potential, $\omega_0=(\omega_x \omega_y \omega_z)^{1/3}$ the geometric trap frequency and $b=(2/ 15^2\pi)^{1/10}/\sqrt{2}\approx 1/\sqrt{7}$\,\cite{books_Pethick_2002}. The 3D Gaussian density distribution normalized to 1,
\begin{equation}
\label{eqA_Gaussian_distribution}
   \left|\Phi^{\mathrm{G}}_\pm(\mathbf{r},t)\right|^2 = \dfrac{1}{(2\pi)^{3/2}\sigma_x\sigma_y\sigma_z}\,\mathrm{Exp}\left[-\dfrac{x^2}{2\sigma_x^2}-\dfrac{y^2}{2\sigma_y^2}-\dfrac{(z\mp z_0)^2}{2\sigma_z^2}\right],
\end{equation}
with $\pm z_0$ the position of the $\pm$ momentum mode. After a direct integration we find 
\begin{equation}
    \chi_{\mathrm{max}}=\dfrac{g}{\hbar}\int_{-\infty}^{\infty}\,d\mathbf{r}\,\left|\Phi^{\mathrm{G}}_{\pm}(\mathbf{r},t)\right|^4=\dfrac{1}{(4\pi\,b^2)^{3/2}\,\hbar}\dfrac{g}{R_x R_y R_z}
\end{equation}
In the case of high atom number the density distribution of the spatial mode can be described by the 3D Thomas-Fermi distribution\,\cite{books_Pethick_2002},
\begin{equation}
    \left|\Phi^{\mathrm{TF}}_{\pm}(\mathbf{r})\right|^2=\dfrac{\mu}{N\,g}\,\left(1-\dfrac{x^2}{R_x^2}-\dfrac{y^2}{R_y^2}-\dfrac{(z\mp z_0)^2}{R_z^2}\right),
\end{equation}
in the region where the right hand side is positive and zero otherwise. A direct integration leads to
\begin{equation}
   \chi_{\mathrm{max}}=\dfrac{g}{\hbar}\int_{-\infty}^{\infty}\,d\mathbf{r}\,\left|\Phi^{\mathrm{TF}}_{\pm}(\mathbf{r},t)\right|^4=\dfrac{15}{14\pi\,\hbar}\dfrac{g}{R_x R_y R_z},
\end{equation}
One can notice that both approaches lead to the same scaling and differ only by $8\%$.

In the case of a variational calculation approach (Eq.\ref{eqA_Gaussian_distribution}), the total non-linear coefficient, $\chi(t)=\chi_S(S)-2\chi_C(t)$ at time $t$ read as, 
\begin{equation}
\label{eq_chi_gaussian}
    \chi(t)=\chi_\mathrm{max}\,\left(1-2\,\mathrm{Exp}\left[-\dfrac{z_0^2}{\sigma_z^2}\right]\right).
\end{equation}
The accumulated non-linear coefficient responsible to the one-axis twisting dynamic at time $t'$ is given by
\begin{equation}
\label{eqA_tau_general}
    \tau_m=\int_0^{mT}\chi(t)\, dt.
\end{equation}
For a total time of $t_{\mathrm{tot}}=m\pi/\omega_z$, with $m$  the number of back and force oscillation of the two spatial mode in the trap, the maximum accumulated non-linear coefficient read
\begin{equation}
\label{eqA_tau_max}
    \tau_{\mathrm{m}}^S=\dfrac{2\, m\pi}{7}\left(15\,a\,\gamma^2\sqrt{\dfrac{m}{\hbar}}\right)^{2/5}\left(\dfrac{\omega_z}{N^3}\right)^{1/5}
\end{equation}
with $\gamma$ the trap ratio: $\gamma = \omega_{x,y}/\omega_z$.

\subsection{Interferometer sequence}

For further clarity, we detail and list here the operations of the non-linear interferometer:
\begin{enumerate}

\item The coherent spin state is generated at $t=0$ by a first $\pi/2$ Bragg pulse where the phase of the laser is fixed to $\phi_L=0$ with an amplitude of $\int\,dt\,\Omega_R\delta(t)=\pi/2$. This step can be described by $\pi/2$ rotation around $\hat{S}_y$.

\item The back and forth oscillation of the two spatial mode between $t=0$ and $t=mT$  generated entanglement though the ``one-axis-twisting'' dynamics. 

\item The orientation of the spin-squeezed state is made possible by changing the laser phase to $\phi_L=0$ with an amplitude of $\int\,dt\,\Omega_R\delta(t)=\alpha$. This step can be described by $\alpha$ rotation around $\hat{S}_x$.

\item The interferometer sequence consists then on two consecutive $\pi/2$-pulse at time $t=mT$ and $t=mT+\tilde{T}/2$ with $\phi_L=\pi/2$ and ($\pi/2$ rotation around $\hat{S}_x$). 

\item A final pulse at $t=mT+\tilde{T}/2$ optimized the rotation of the output state as described step 3 with $\int\,dt\,\Omega_R\delta(t)=\beta$.

\end{enumerate}

\section{Calculation of the gain of the interferometer.}
\label{sec_Appendix_B}
\subsection{Input state preparation and linear interferometry sequence.}
\label{sec_Appendix_B1}
We consider the initial coherent spin state (CSS)
\begin{equation}
\label{eq_CSS_input}
    |\psi_0\rangle = 2^{-S}\sum \limits_{n=0}^{2S}\binom{2S}{n}^{1/2}|S,S-n\rangle,
\end{equation}
where $S=N/2$ and we define $|\psi_e\rangle=\hat{U}|\psi_0\rangle$ with  $\hat{U}= e^{-i\tau \hat{S}_z^2}$. The mean value of a generic operator $\hat{Q}$ at the input port of the interferometer read as 
\begin{equation}
    \langle\,\hat{Q}\,\rangle_\alpha = \langle \psi_e| \,\hat{Q}_\alpha\,|\psi_e\rangle
\end{equation}
where $\hat{Q}_\alpha= R_x^\dagger(\alpha)\,\hat{Q}\,R_x(\alpha)$ and $R_x(\alpha)=e^{-i\alpha\hat{S}_x}$. We recall the relation, 
\begin{equation}
\label{eq_App_RotationSy_and_Sz}
    R_x^\dagger(\alpha)\,S_{y/z}\,R_x(\alpha)=\cos(\alpha)\,S_{y/z}\mp\sin(\alpha)\,S_{z/y}.
\end{equation}
In order to evaluate the gain at the input of the interferometer, $\emph{i.e.}$ the spin-squeezing parameter introduced the main text, we have to evaluate the pseudo-spin operators $\{\hat{S}_x,\hat{S}_y,\hat{S}_z\}$ in addition to $\{\hat{S}_x^2,\hat{S}_y^2,\hat{S}_z^2\}$ where the operators can be expressed with respect to the raising and lowering operators $S_{\pm}$: $S_x=(S_++S_-)/2$, $S_y=(S_+-S_-)/2i$, $S_z=(S_+S_- -S_-S_+)/2$, $ \hat{S}_{x/y}^2=\left(\pm\hat{S}_+^2+\pm\hat{S}_-^2+2S_+S_--2S_z\right)/4$, $\hat{S}_z^2=S^2-S_x^2-S_y^2$.  

Using the relations $\hat{U}^\dagger S_{+} \hat{U} = S_{+} e^{i \tau ( 2 S_z +1) }$ and $\hat{S}_-=\left[\hat{S}_{+}\right]^\dagger$ it is convenient to calculate the following list:
\begin{subequations}
\label{eq_Appendix_Sp_Sm}
\begin{align}
    &\langle\psi_e|\,\hat{S}_+|\psi_e\rangle\equiv\langle\psi_0|\,\hat{S}_+\,e^{2i\tau(\hat{S}_z+\frac{1}{2})}|\psi_0\rangle=S\,\cos\left(\tau\right)^{2S-1}\\
    &\langle\psi_e|\,\hat{S}_+^2|\psi_e\rangle\equiv\langle\psi_0|\,\hat{S}_+^2\,e^{4i\tau(\hat{S}_z+1)}|\psi_0\rangle=S\left(S-\frac{1}{2}\right)\cos(2\tau)^{2S-2}\\
    &\langle\psi_e|\,\hat{S}_+^3|\psi_e\rangle\equiv\langle\psi_0|\,\hat{S}_+^3\,e^{6i\tau(\hat{S}_z+\frac{3}{2})}|\psi_0\rangle=S\left(S-\frac{1}{2}\right)(S-1)\cos(3\tau)^{2S-3}\\
    &\langle\psi_e|\,\hat{S}_+\hat{S}_z|\psi_e\rangle\equiv\langle\psi_0|\,\hat{S}_+\hat{S}_z\,e^{2i\tau(\hat{S}_z+\frac{1}{2})}|\psi_0\rangle=iS\left(S-\frac{1}{2}\right)\cos\left(\tau\right)^{2S-2}\sin\left(\tau\right)-\frac{S}{2}\cos\left(\tau\right)^{2S-1}\\
    &\langle\psi_e|\,\hat{S}_+^2\hat{S}_z|\psi_e\rangle\equiv\langle\psi_0|\,\hat{S}_+^2\hat{S}_z\,e^{4i\tau(\hat{S}_z+1)}|\psi_0\rangle=iS\left(S-\frac{1}{2}\right)(S-1)\cos\left(2\tau\right)^{2S-3}\sin\left(2\tau\right)-S\left(S-\frac{1}{2}\right)\cos\left(2\tau\right)^{2S-2}\\
    &\langle\psi_e|\,\hat{S}_+\hat{S}_z^2|\psi_e\rangle\equiv\langle\psi_0|\,\hat{S}_+\hat{S}_z^2\,e^{2i\tau(\hat{S}_z+1)}|\psi_0\rangle=\dfrac{S}{4}\cos(\tau)^{2S-1}+\dfrac{S}{2}\left(S-\dfrac{1}{2}\right)\cos(\tau)^{2S-1} \nonumber\\
    &-i S\left(S-\dfrac{1}{2}\right) \cos(\tau)^{2S-2}\sin(\tau)-S\left(S-\dfrac{1}{2}\right)(S-1)\cos(\tau)^{2S-3}\sin(\tau)^2 \\
    &\langle\psi_e|\,\hat{S}_+\hat{S}_-|\psi_e\rangle\equiv\langle\psi_0|\,\hat{S}_+\hat{S}_-|\psi_0\rangle=S\left(S+\frac{1}{2}\right), \\
    &\langle\psi_e|\,\hat{S}_+^2\hat{S}_-|\psi_e\rangle\equiv\langle\psi_0|\,\hat{S}_+\hat{S}_-|\psi_0\rangle=S\left(S-\frac{1}{2}\right)(S-1)\cos(\tau)^{2S-3}+2S\left(S-\frac{1}{2}\right)\cos(\tau)^{2S-2}e^{i\tau}, \\
    &\langle\psi_e|\,\hat{S}_z|\psi_e\rangle\equiv\langle\psi_0|\,\hat{S}_z|\psi_0\rangle=0, \\
    &\langle\psi_e|\,\hat{S}_z^2|\psi_e\rangle\equiv\langle\psi_0|\,\hat{S}_z^2|\psi_0\rangle=\frac{S}{2}.
\end{align}
\end{subequations}
Using Eq.\,(\ref{eq_App_RotationSy_and_Sz}), the different operators are then given by\,\cite{PRA_Kitagawa_1993}:
\begin{subequations}
\label{eq_App_Jx_Jy_Jz_LAI}
\begin{align}
\label{eq_Jx_LAI}
\langle\,S_x\,\rangle_\alpha&=S\cos(\tau)^{2S-1}, \\
\label{eq_Jy_Jz_LAI}
\langle\,S_{y/z}\,\rangle_\alpha&=0, \\
\label{eq_Jx2_LAI}
\langle\,S_x^2\,\rangle_\alpha&=\dfrac{S}{2}\left[2S-\left(S-\dfrac{1}{2}\right)A\right] \\
\label{eq_Jy2_Jz2_LAI}
\langle\,S_{y/z}^2\,\rangle_\alpha &= \dfrac{S}{2}\left(1+\dfrac{2S-1}{4}\left[A\pm \sqrt{A^2+B^2}\cos(2\alpha+2\delta)\right]\right)
\end{align}
\end{subequations}
with $A=1-\cos(2\tau)^{2S-2}$, $B=4\sin(\tau)\cos(\tau)^{2S-2}$ and  $\delta=\mathrm{arctan}\left(B/A\right)/2$. The spin-squeezing parameter therefore read as:
\be
\xi^2=\dfrac{4+(2S-1)(A-\sqrt{A^2+B^2})}{4\cos(\tau)^{4S-2}},
\ee
and lead in the case of a linear atom interferometer sequence to the gain $\mathcal{G}_{\rm L}=1/\xi$.

\subsection{Non-linear interferometry sequence}
\label{sec_Appendix_B2}

\subsubsection{The linear error propagation}

The sensitivity of the interferometer, $\Delta \theta$, is evaluated via the linear error propagation formula:
\be
\label{eq_App_ErrorPropagator}
(\Delta \theta)^2 = \frac{(\Delta S_z)^2_{\rm out}}{( d\langle \hat{S}_z \rangle_{\rm out} / d \theta )^2},
\ee
where the state at the end of the interferometer sequence is given by:
\be
|\Psi_{\rm out}\rangle=e^{i\pi/2\hat{S}_x}e^{-i\beta\hat{S}_x}e^{-i\tilde{\tau}\hat{S}_z^2}e^{-i\theta\hat{S}_z}e^{-i\pi/2\hat{S}_x}e^{-i\alpha\hat{S}_x}|\Psi_e\rangle\equiv e^{-i\beta\hat{S}_x}e^{-i\tilde{\tau}\hat{S}_y^2}e^{-i\theta\hat{S}_y}e^{-i\alpha\hat{S}_x}|\Psi_e\rangle.
\ee
In this case, $\langle \hat{S}_z \rangle_{\rm out}= \langle \Psi_e| e^{i\alpha\hat{S}_x} e^{i\theta\hat{S}_y} e^{i\tilde{\tau}\hat{S}_y^2} e^{i\beta\hat{S}_x} \hat{S}_z e^{-i\beta\hat{S}_x}e^{-i\tilde{\tau}\hat{S}_y^2}e^{-i\theta\hat{S}_y}e^{-i\alpha\hat{S}_x}|\Psi_e\rangle$ can be simplified to 
\be
\langle \hat{S}_z \rangle_{\rm out}= \langle \Psi_e| e^{i\alpha\hat{S}_x} e^{i\tilde{\tau}\hat{S}_y^2} \left(\cos(\theta)\hat{S}_z-\sin(\theta)\hat{S}_x\right) e^{-i\tilde{\tau}\hat{S}_y^2}e^{-i\alpha\hat{S}_x}|\Psi_e\rangle \cos(\beta)+ \langle \Psi_e| e^{i\alpha\hat{S}_x} \hat{S}_y e^{-i\alpha\hat{S}_x}|\Psi_e\rangle \sin(\beta),
\ee
where we have used the relations, $e^{i\beta\hat{S}_x} \hat{S}_z e^{-i\beta\hat{S}_x}=\cos(\beta)\hat{S}_z-\sin(\beta)\hat{S}_y$ and $e^{i\theta\hat{S}_y} \hat{S}_z e^{-i\theta\hat{S}_y}=\cos(\theta)\hat{S}_z-\sin(\theta)\hat{S}_x$. The denominator of Eq.\ref{eq_App_ErrorPropagator} read then as
\be
\dfrac{d}{d \theta}\langle \hat{S}_z \rangle_{\rm out} = -\langle \Psi_e| e^{i\alpha\hat{S}_x} e^{i\tilde{\tau}\hat{S}_y^2} \left(\sin(\theta)\hat{S}_z+\cos(\theta)\hat{S}_x\right) e^{-i\tilde{\tau}\hat{S}_y^2}e^{-i\alpha\hat{S}_x}|\Psi_e\rangle \cos(\beta).
\ee
In the particular case $\theta=0$, we find
\be
\label{eq_app_denominator}
\dfrac{d}{d \theta}\langle \hat{S}_z \rangle_{\rm out}\Big|_{\theta=0} = -\langle \Psi_e| e^{i\alpha\hat{S}_x} e^{i\tilde{\tau}\hat{S}_y^2} \hat{S}_x e^{-i\tilde{\tau}\hat{S}_y^2}e^{-i\alpha\hat{S}_x}|\Psi_e\rangle \cos(\beta)\equiv -\cos(\beta) \langle \hat{S}_x \rangle_{\rm out},
\ee
where we have used the relation: $e^{i\beta\hat{S}_x} \hat{S}_x e^{-i\beta\hat{S}_x}=\hat{S}_x$. Frm Eqs.~\ref{eq_App_ErrorPropagator} and~\ref{eq_app_denominator} we define the sensitivity gain as $\Delta \theta=1/\sqrt{N} \mathcal{G}_{\alpha,\beta}$, with 
\be
\label{eq_app_GainNL}
\left(\mathcal{G}_{\alpha,\beta}\right)^2=\dfrac{\cos^2(\beta)\langle \hat{S}_x\rangle^2_{\rm out}}{N (\Delta \hat{S}_z)^2_{\rm out}}.
\ee

\subsubsection{Calculation of the different part}

In this section we want to calculate the gain at the output port of the interferometer sequence in the case where $\tilde{\tau}\neq 0$. We want to evaluate the quantity $\langle\,\hat{Q}\,\rangle_{\alpha,\beta}\equiv\langle\Psi_{\rm out}|\,\hat{Q}\,|\Psi_{\rm out}\rangle_{\alpha,\beta}$. In the case where $\theta$ and $\tilde{\tau}$ are considered small we have
\begin{equation}
\label{App_Q_out}
    \langle\,\hat{Q}\,\rangle_{\alpha,\beta} = \langle \psi_e| \hat{Q}_{\alpha,\beta}|\psi_e\rangle 
    + i\theta\, \langle \psi_e|[R_x^\dagger(\alpha) \hat{S}_y R_x(\alpha),\hat{Q}_{\alpha,\beta}] |\psi_e\rangle
    + i\tilde{\tau}\, \langle \psi_e|[R_x^\dagger(\alpha) \hat{S}_y^2 R_x(\alpha),\hat{Q}_{\alpha,\beta}] |\psi_e\rangle. 
\end{equation}
with $\hat{Q}$ being $\{\hat{S}_x,\hat{S}^2_z\}$ and $\hat{Q}_{\alpha,\beta}=R_x^\dagger(\alpha+\beta)\hat{Q}R_x(\alpha+\beta)$. Using Eq.\,(\ref{eq_App_RotationSy_and_Sz}) we have to calculate
\begin{subequations}
\begin{align}
    \langle \hat{S}_x\rangle_{\rm \alpha,\beta}&=\langle \psi_e| \hat{S}_x|\psi_e\rangle\\
    &+i\theta\langle \psi_e| \left[\cos(\alpha)\hat{S}_y-\sin(\alpha)\hat{S}_z,\hat{S}_x\right]|\psi_e\rangle \nonumber \\
    &+i\tilde{\tau}\langle \psi_e| \left[\cos^2(\alpha)\hat{S}_y^2+\sin^2(\alpha)\hat{S}_z^2-\frac{\sin(2\alpha)}{2}(\hat{S}_y\hat{S}_z+\hat{S}_z\hat{S}_y),\hat{S}_x\right]|\psi_e\rangle, \nonumber
\end{align}    
\end{subequations}
and
\begin{subequations}
\begin{align}
    &\langle \hat{S}_z^2\rangle_{\rm \alpha,\beta}=\langle \psi_e| \cos^2(\alpha+\beta)\hat{S}_z^2+\sin^2(\alpha+\beta)\hat{S}_y^2+\frac{\sin(2\alpha+2\beta)}{2}(\hat{S}_y\hat{S}_z+\hat{S}_z\hat{S}_y)|\psi_e\rangle \\
    &+i\theta\langle \psi_e| \left[\cos(\alpha)\hat{S}_y-\sin(\alpha)\hat{S}_z,\cos^2(\alpha+\beta)\hat{S}_z^2+\sin^2(\alpha+\beta)\hat{S}_y^2+\frac{\sin(2\alpha+2\beta)}{2}(\hat{S}_y\hat{S}_z+\hat{S}_z\hat{S}_y)\right]|\psi_e\rangle \nonumber \\
    &+i\tilde{\tau}\langle \psi_e| \left[\cos^2(\alpha)\hat{S}_y^2+\sin^2(\alpha)\hat{S}_z^2-\frac{\sin(2\alpha)}{2}(\hat{S}_y\hat{S}_z+\hat{S}_z\hat{S}_y),\cos^2(\alpha+\beta)\hat{S}_z^2+\sin^2(\alpha+\beta)\hat{S}_y^2+\frac{\sin(2\alpha+2\beta)}{2}(\hat{S}_y\hat{S}_z+\hat{S}_z\hat{S}_y)\right]|\psi_e\rangle. \nonumber
\end{align}    
\end{subequations}
The commutators appearing in the above equation are calculated using
\begin{subequations}
\label{eq_Appendix_Commutateurs}
\begin{align}
    &\left[\hat{S}_y,\hat{S}_x\right]=-i\hat{S}_z\\
    &\left[\hat{S}_z,\hat{S}_x\right]=i\hat{S}_y\\
    &\left[\hat{S}_y^2,\hat{S}_x\right]=\left(\hat{S}_z+\frac{1}{2}\right)\hat{S}_--\hat{S}_+\left(\hat{S}_z+\frac{1}{2}\right)\\
    &\left[\hat{S}_z^2,\hat{S}_x\right]=\hat{S}_+\left(\hat{S}_z+\frac{1}{2}\right)-\left(\hat{S}_z+\frac{1}{2}\right)\hat{S}_-\\
    &\hat{S}_y\hat{S}_z+\hat{S}_z\hat{S}_y=i\left(\hat{S}_z+\frac{1}{2}\right)\hat{S}_--i\hat{S}_+\left(\hat{S}_z+\frac{1}{2}\right) \\
    &\left[\hat{S}_y\hat{S}_z+\hat{S}_z\hat{S}_y,\hat{S}_x\right]=\frac{1}{2i}\left(\hat{S}_+^2+\hat{S}_-^2+4\hat{S}_z^2+2\hat{S}_z-2\hat{S}_+\hat{S}_-\right)
\end{align}
\end{subequations}
and
\begin{subequations}
\label{eq_Appendix_Commutateurs_2}
\begin{align}
    &\left[\hat{S}_y,\hat{S}_z^2\right]=i\hat{S}_+\left(\hat{S}_z+\frac{1}{2}\right)+i\left(\hat{S}_z+\frac{1}{2}\right)\hat{S}_-\\
    &\left[\hat{S}_z,\hat{S}_y^2\right]=\frac{\hat{S}_-^2-\hat{S}_+^2}{2} \\
    &\left[\hat{S}_y,\hat{S}_y\hat{S}_z+\hat{S}_z\hat{S}_y\right]=\frac{\hat{S}_+^2-\hat{S}_-^2}{2} \\
    &\left[\hat{S}_z,\hat{S}_y\hat{S}_z+\hat{S}_z\hat{S}_y\right]=-i\hat{S}_+\left(\hat{S}_z+\frac{1}{2}\right)-i\left(\hat{S}_z+\frac{1}{2}\right)\hat{S}_-\\
    &\left[\hat{S}_y^2,\hat{S}_z^2\right]=\hat{S}_+^2(\hat{S}_z+1)-(\hat{S}_z+1)\hat{S}_-^2 \\
    &\left[\hat{S}_y\hat{S}_z+\hat{S}_z\hat{S}_y,\hat{S}_z^2\right]=-\frac{1}{2i}\left(4\hat{S}_+\hat{S}_z^2+4\hat{S}_+\hat{S}_z+\hat{S}_++4\hat{S}_z^2\hat{S}_-+4\hat{S}_z\hat{S}_-+\hat{S}_-\right)\\
    &\left[\hat{S}_y\hat{S}_z+\hat{S}_z\hat{S}_y,\hat{S}_y^2\right]=\frac{1}{2i}\left( \hat{S}_+^2\hat{S}_-+\hat{S}_+\hat{S}_-^2-2(\hat{S}_+\hat{S}_z+\hat{S}_z\hat{S}_-)-\hat{S}_+^3-\hat{S}_-^3-\hat{S}_+-\hat{S}_-\right).
\end{align}
\end{subequations}

\subsubsection{Evaluation of the sensitivity gain in the perturbative regime}

For safe of simplicity it is convenient to introduce the notation $\langle\hat{X}\rangle_{\alpha,\beta}=\langle\hat{X}^{(0)}\rangle_{\alpha,\beta}+\tilde{\tau}\langle\hat{X}^{(1)}\rangle_{\alpha,\beta}$. Equation~\ref{eq_app_GainNL} becomes 
\be
\left(\mathcal{G}_{\alpha,\beta}\right)^2=\dfrac{\cos^2(\beta)}{N}\dfrac{\left(\langle\hat{S}_x^{(0)}\rangle_{\alpha,\beta}+\tilde{\tau}\langle\hat{S}_x^{(1)}\rangle_{\alpha,\beta}\right)^2}{ \langle\hat{S}_z^{2\,(0)}\rangle_{\alpha,\beta}+\tilde{\tau}\langle\hat{S}_z^{2\,(1)}\rangle_{\alpha,\beta}}\equiv \cos^2(\beta)\dfrac{\langle\,\hat{S}_{x}^{(0)}\,\rangle^2_{\alpha,\beta}}{N\langle\,\hat{S}_{z}^{2\,(0)}\,\rangle_{\alpha,\beta}}\left(1+2\tilde{\tau}\dfrac{\langle\,\hat{S}_{x}^{(1)}\,\rangle_{\alpha,\beta}}{\langle\,\hat{S}_{x}^{(0)}\,\rangle_{\alpha,\beta}}-\tilde{\tau}\dfrac{\langle\,\hat{S}_{z}^{2\,(1)}\,\rangle_{\alpha,\beta}}{\langle\,\hat{S}_z^{2\,(0)}\,\rangle_{\alpha,\beta}}\right),
\ee
where $\langle\,\hat{S}_{x}^{(0)}\,\rangle^2_{\alpha,\beta}/\langle\,\hat{S}_{z}^{2\,(0)}\,\rangle_{\alpha,\beta}\equiv 1/\xi^2$ is the spin-squeezing parameter introduced in\,\cite{Nature_Sorensen_2001}. After calculation we find,
\begin{subequations}
\label{eq_Appendix_Sx_alpha}
\begin{align}
\langle\,\hat{S}_{x}^{(0)}\,\rangle_{\alpha,\beta} &=S\cos(\tau)^{2S-1}, \\
\langle\,\hat{S}_{x}^{(1)}\,\rangle_{\alpha,\beta} &=S\left(S-\dfrac{1}{2}\right)\left[2 \cos(\tau)^{2S-2}\sin(\tau)\cos(2\alpha)+(1-\cos(2\tau)^{2S-2})\dfrac{\sin(2\alpha)}{2}\right], \\
\langle\,\hat{S}_{z}^{2\,(0)}\,\rangle_{\alpha,\beta} &=\dfrac{S}{2}\cos^2(\alpha+\beta)+S(2S-1)\cos(\tau)^{2S-2}\sin(\tau)\dfrac{\sin(2\alpha+2\beta)}{2}\\ \nonumber
&+S\left[1+2S-(2S-1)\cos(2\tau)^{2S-2}\right]\dfrac{\sin(\alpha+\beta)^2}{4}, \\
\langle\,\hat{S}_{z}^{2\,(1)}\,\rangle_{\alpha,\beta} &=-S\left(S-\dfrac{1}{2}\right)\bigg( \bigg. 2(S-1)\cos(2\tau)^{2S-3}\sin(2\tau)\cos(2\alpha+\beta)-(S-1)\cos(3\tau)^{2S-3}\cos(\alpha)\sin(\alpha+\beta)  \nonumber \\
&+\cos(\tau)^{2S-3}\left[-2\cos(\alpha+\beta)(\cos(\tau)^2-2(S-1)\sin(\tau)^2)\sin(\alpha)+(S+\cos(2\tau)\cos(\alpha)\sin(\alpha+\beta) ) \right]\bigg. \bigg).
\end{align}
\end{subequations}
To first order in $\tau$ we have:
\begin{subequations}
\begin{align}
\dfrac{\langle\,\hat{S}_{x}^{(0)}\,\rangle^2_{\alpha,\beta}}{N\langle\,\hat{S}_{z}^{2\,(0)}\,\rangle_{\alpha,\beta}}&=1-(2S-1)\sin(2\alpha+2\beta)\tau,\\
\dfrac{\langle\,\hat{S}_{x}^{(1)}\,\rangle_{\alpha,\beta}}{\langle\,\hat{S}_{x}^{(0)}\,\rangle_{\alpha,\beta}}&=(2S-1)\cos(2\alpha)\tau, \\
\dfrac{\langle\,\hat{S}_{z}^{2\,(1)}\,\rangle_{\alpha,\beta}}{\langle\,\hat{S}_{z}^{2\,(0)}\,\rangle_{\alpha,\beta}}&=-(2S-1)\left[\sin(2\beta)+\cos(\beta)\{(2S-3)\cos(2\alpha+\beta)+(2S-1)\cos(2\alpha+3\beta)\}\tau\right].
\end{align}
\end{subequations}
The sensitivity gain therefore read to first order in $\tau$ and $\tilde{\tau}$ as
\be
\left(\mathcal{G}_{\alpha,\beta}\right)^2=\left[1+(2S-1) \{\sin(2\beta)\tilde\tau-\sin(2\alpha+2\beta)\tau\}\right]\cos^2(\beta).
\ee
In the case where $\alpha=0$ we have:
\be
\left(\mathcal{G}_{0,\beta}\right)^2=\left[1+(2S-1)\sin(2\beta) (\tilde\tau-\tau)\right]\cos^2(\beta).
\ee
In the case where $\alpha=\pi/2$ we have:
\be
\left(\mathcal{G}_{\pi/2,\beta}\right)^2=\left[1+(2S-1)\sin(2\beta) (\tau+\tilde\tau)\right]\cos^2(\beta).
\ee

\twocolumngrid


\begin{thebibliography}{10}
\bibitem{data_description}
In our simulations, we specialize to the case of a BEC of Rubidium 87 with
  $N=10^3$ atoms. The s-wave scattering length is fix to
  $a=99.6\,a_0$\,\cite{dataRb} with $a_0$ being the Bohr radius, and we defined
  the wave vector $k_L=2\pi/\lambda_L$ of the laser with $\lambda_L=852$\,nm.
\end{thebibliography}

\end{document}